\documentclass[11pt]{article}
\usepackage[utf8]{inputenc}
\usepackage[T1]{fontenc}
\usepackage[margin=1in]{geometry}
\usepackage{amsmath,amssymb}
\usepackage{graphicx}
\usepackage{booktabs}
\usepackage{array}
\usepackage{adjustbox}
\usepackage{placeins}
\usepackage{lmodern}
\usepackage{microtype}
\usepackage{caption}
\usepackage[affil-it]{authblk}
\usepackage{url}
\usepackage{textcomp}
\usepackage[colorlinks=true,linkcolor=blue,citecolor=blue,urlcolor=blue]{hyperref}
\providecommand{\tightlist}{\setlength{\itemsep}{0pt}\setlength{\parskip}{0pt}}
\title{\textbf{Treating the Model as an Untrusted Component: Deterministic Policy Enforcement, Human Approval and Taint Tracking Against Prompt Injection in Tool-Using Agents, with Validation on Open-Weights Models}}

\author[1,*]{Mustafa S. Aljumaily}
\author[2]{Nawar S. Alseelawi}
\author[3]{Raghad Alshabandar}
\affil[1]{ICT Department, Missan Oil Company, Missan, Iraq\\ \texttt{mustafasadiq0@gmail.com}; ORCID 0000-0002-7886-9537}
\affil[2]{University of Misan, Maysan, Iraq\\ \texttt{nawar.alseelawi@uomisan.edu.iq}; ORCID 0000-0001-5739-2349}
\affil[3]{National Center for Artificial Intelligence, Baghdad, Iraq\\ \texttt{raghadalshabandar@outlook.com}; ORCID 0000-0002-1393-3761}
\affil[*]{Corresponding author: \texttt{mustafasadiq0@gmail.com}}
\date{}

\begin{document}

\maketitle

\begin{abstract}
\noindent
Tool-using LLM agents act on instructions embedded in the data they read, so prompt injection is a structural rather than a prompt-wording problem. We compare in-prompt defences (hardened prompts, spotlighting, LLM detectors) with deterministic enforcement outside the model (four reference monitors of increasing strength), human approval and taint tracking, judged by one effect-based oracle independent of every defence in an executable environment of 13 tasks and 12 tools, with argument-fuzzing attackers and severity-weighted harm. In an evaluation with real open-weights models (Qwen2.5-1.5B and Qwen2.5-7B; 156 attack episodes per defence and model), the models proposed the attacker's call in 64\% and 84\% of attack episodes behind the task-bound monitor, which admitted 1.3\% and 7.7\%; guardrail effects depended on the model (a prompt guardrail moved obedience from 67\% to 65\% and from 86\% to 55\%, spotlighting plus a detector to 12\% and 0\%) against static attacks. Simulation with an assumed model covers what no real run did: under the assumed erosion law a persistent adaptive attacker or a compromised model drives every guardrail stack to 100.0\% attack success, whereas the task-bound monitor holds severe actions at 0.0\%; its residual (harm 0.69/10) falls by 62\% under taint-triggered human review, and a human layer without a deterministic monitor fails against approval spamming unless each episode has a deterministic approval budget. Implementation quality matters: a naive monitor admits 18 of 33 effective attack variants, a hardened one 6 and a capability-bound one 1. The real-model evidence is limited to two open-weights models and static attacks, the human-approval and taint results are simulation-only, and the larger model's detector stack contradicts the assumed erosion law. Code, data, a two-rater oracle audit and a notebook for repeating the real-model evaluation are provided.

\medskip
\noindent\textbf{Keywords:} prompt injection, LLM agents, reference monitor, capability security, taint tracking, human-in-the-loop, adaptive attacks, zero trust, simulation.
\end{abstract}

\hypertarget{introduction}{%
\FloatBarrier
\section{Introduction}\label{introduction}}

An agent that reads email, browses pages and queries databases, and then sends messages, edits configuration or sets an actuator value, mixes instructions and data in one channel \cite{greshake,zverev}. The prevailing mitigation makes the model better at ignoring injected text; its security argument is probabilistic and assumes a non-adaptive attacker \cite{nasr,zhan25}. The alternative is classical: treat the model as an untrusted, non-deterministic component and let a small deterministic mechanism authorise every proposed action \cite{anderson,saltzer}. A pragmatic third option keeps a human in the loop for high-impact actions, but humans fatigue and can be spammed \cite{alahmadi}. The question matters most in industrial and critical-infrastructure deployments, where an agent may hold tools that change configuration or actuator setpoints and a single unauthorised action can have physical consequences \cite{nist80082}. This paper compares the three approaches under one effect-based oracle and one attacker model, and asks which guarantees survive a persistent, adaptive or fully compromised model. Sections 3-5 define the threat model, the defences and the methodology; Section 6 reports simulation and real-model results; Sections 7-9 discuss implications, limits and availability.

\textbf{Contributions.} (1) A reproducible executable harness with an effect-based oracle and a ladder of four monitors implemented naively and hardened. (2) A severity-weighted harm metric, showing that binary attack-success rate hides which actions an attacker obtains. (3) A human-approval and taint-tracking layer with reviewer fatigue, approval spamming and a deterministic approval budget. (4) Failure analyses of the taint tracker and of policy completeness, each crossed with the layering. (5) A measurement and calibration pipeline that tests the susceptibility law on a real open-weights model (the assumed monotone law is rejected). (6) A consistency check against a separately built abstract model (Section 6.9; the two share design assumptions, so this is not independent evidence). (7) An end-to-end evaluation with real open-weights models running the same tasks, oracle and monitors (Section 6.11), and a notebook to repeat it at scale.

\textbf{Hypotheses.} Hyp. 1 to Hyp. 5 were fixed before the executable-harness experiments were run; Hyp. 6 and Hyp. 7 were specified for the extension study. Hyp. 1: deterministic enforcement is invariant to model susceptibility; guardrails are monotone in it. Hyp. 2: under adaptive, repeated attack guardrail success tends to 1 while a sound monitor plateaus at its policy residual. Hyp. 3: monitor implementation quality matters against attackers that fuzz arguments. Hyp. 4: a sound monitor's residual is dominated by in-policy semantic manipulation. Hyp. 5: deterministic enforcement costs utility mainly through policy incompleteness. Hyp. 6: a human layer helps only behind a deterministic monitor, and over-escalation degrades it through fatigue. Hyp. 7: a deterministic approval budget restores a bound against approval spamming.

\hypertarget{related-work}{%
\FloatBarrier
\section{Related work}\label{related-work}}

\textbf{Attacks, benchmarks and guidance.} Indirect prompt injection was characterised for LLM-integrated applications by Greshake et al.~\cite{greshake}, following the direct goal-hijacking and prompt-leaking attacks on GPT-3 of Perez and Ribeiro \cite{perez}. The threat is now catalogued by the NIST adversarial machine-learning taxonomy \cite{nist100} and by the OWASP list of risks specific to agentic applications \cite{owasp}. Benchmarks measure it on tool-using agents: InjecAgent \cite{injecagent} provides 1,054 test cases, AgentDojo \cite{agentdojo} 97 tasks and 629 security test cases, ASB \cite{asb} a broader attack-and-defence suite, WASP \cite{wasp} realistic web-agent hijacking objectives and AgentDyn \cite{agentdyn} dynamic, open-ended tasks; Liu et al.~\cite{liu24} formalise attacks and defences and benchmark five attacks and ten defences across ten LLMs.

\textbf{In-band defences and adaptive attackers.} Prompt-level defences ask the model itself to resist: instruction hierarchies \cite{wallace} train models to ignore lower-privileged instructions, spotlighting \cite{hines} marks the provenance of untrusted input and was reported to cut attack success on GPT-family models from above 50\% to below 2\% against a static attack set, StruQ \cite{struq} separates prompt and data with structured queries and a fine-tuned model, and SecAlign \cite{secalign} trains the same separation by preference optimisation. Whether models can separate instructions from data at all is contested: Zverev et al.~\cite{zverev} introduce a formal measure of separation and find that the evaluated models separate them poorly. Adaptive attackers undermine such defences: Nasr et al.~\cite{nasr} bypass 12 recent defences with success above 90\% for most of them although most had reported near-zero rates, Zhan et al.~\cite{zhan25} bypass eight defences against indirect injection on agents with success above 50\%, and Jia et al.~\cite{jia25} show that the fine-tuning defences StruQ and SecAlign fall to adaptive attacks and lose utility, and that detectors judged by AUC alone are misleading. This follows the general lesson of the adversarial-example literature that a defence must be evaluated against attacks designed for it \cite{tramer} and motivates our adaptive and compromised-model attackers.

\textbf{Out-of-band defences.} Architectural defences move the security decision out of the model. CaMeL \cite{camel} extracts control and data flow from the trusted query and enforces capability-based policies at tool calls, solving 77\% of AgentDojo tasks with provable security against 84\% for an undefended system; FIDES \cite{fides} tracks confidentiality and integrity labels in a planner and enforces policies deterministically; Progent \cite{progent} provides programmable privilege control over tool calls; AgentSpec \cite{agentspec} provides customisable runtime enforcement of rules over agent actions; Conseca \cite{conseca} generates context-specific security policies; IsolateGPT \cite{isolategpt} isolates the execution of agentic applications; MELON \cite{melon} detects injection by re-running the agent on a masked user prompt and comparing the tool calls; the Task Shield \cite{taskshield} checks every instruction and tool call for alignment with the user's task; and Beurer-Kellner et al.~\cite{patterns} catalogue six design patterns. A recent preprint \cite{narisetty} re-evaluates an out-of-band defence under an adaptive attack with an open-weights agent. Our study differs in what it holds constant and what it adds: every defence family is judged by one effect-based oracle against one attacker model, argument-level implementation quality and policy completeness are varied explicitly, human-approval and taint-tracking layers are added, and the results are checked on real open-weights models.

\textbf{Foundations.} The conceptual roots are the reference monitor \cite{anderson}, the protection design principles of Saltzer and Schroeder \cite{saltzer}, capability-based addressing \cite{dennis}, the confused-deputy problem \cite{hardy}, which an injected agent re-creates, and lattice-based information flow \cite{denning}. Zero-trust architecture \cite{nist207} and operational-technology security guidance \cite{nist80082} supply the deployment context. Our monitors are deliberately simple instances of this tradition and not re-implementations of any one system.

\hypertarget{threat-model}{%
\FloatBarrier
\section{Threat model}\label{threat-model}}

\emph{Assets.} Read tools (\texttt{fetch\_\allowbreak{}url}, \texttt{read\_\allowbreak{}file}, \texttt{read\_\allowbreak{}email}, \texttt{query\_\allowbreak{}db}, \texttt{list\_\allowbreak{}files}) are the injection channel; side-effecting tools (\texttt{send\_\allowbreak{}email}, \texttt{http\_\allowbreak{}post}, \texttt{write\_\allowbreak{}file}, \texttt{delete\_\allowbreak{}file}, \texttt{update\_\allowbreak{}config}, \texttt{set\_\allowbreak{}setpoint}, \texttt{run\_\allowbreak{}shell}) are the targets. The user's request is the only trusted statement of intent.

\emph{Attackers.} A0 naive; A1 authority-spoofing; A2 obfuscated/delimiter-spoofing; A3 adaptive (strength grows with attempts, guardrail effectiveness erodes, three argument variants per attempt, blocked variants remembered); A4 fully compromised model (obeys with probability 1 and tries the whole attack library with every variant, including approval spamming).

\emph{Trusted base.} Monitor, policy, tool implementations, taint tracker and the approval channel. Out of scope: compromise of those, covert channels, denial of service.

\emph{Objective.} No executed side-effecting call may differ from what the user's request authorises, judged by an oracle that canonicalises arguments and never consults a defence.

\hypertarget{defences-under-test}{%
\FloatBarrier
\section{Defences under test}\label{defences-under-test}}

\begin{table}[htbp]
\centering
\caption{Defences under test.}
\label{tab:1}
\small
\begin{tabular}{l p{0.72\linewidth} l}
\toprule
\textbf{Key} & \textbf{Defence} & \textbf{Deterministic} \\
\midrule
D0 & No defence & no \\
D1 & In-prompt guardrail & no \\
D2 & Guardrail + spotlighting & no \\
D3 & Spotlight + LLM detector & no \\
D4 & P1 tool allow-list & yes \\
D5 & P2n naive argument checks & yes \\
D6 & P2h hardened argument checks & yes \\
D7 & P3 task-bound capabilities + confidentiality flow rule & yes \\
D8 & P3 + spotlight + detector & partly \\
H1 & P3 + human approval for severity $\geq$ 8 & no (human) \\
H2 & P3 + human + taint: when the session is tainted, severity $\geq$ 3 also needs approval & no (human) \\
H3 & P2h + human + taint (no capabilities) & no \\
H4 & P1 + human + taint & no \\
H5 & Human + taint only (no monitor) & no \\
\bottomrule
\end{tabular}
\end{table}

\textbf{Severity.} write\_file 3, send\_email 5, delete\_file 7, http\_post and run\_shell 8, update\_config and set\_setpoint 9; egress carrying confidential data 10. Severity is per tool, not per content, so a semantically false write is rated 3; an operator who rates such content higher should lower \texttt{taint\_\allowbreak{}thr} accordingly.

\textbf{Human approval.} Calls that pass the monitor and have severity \textgreater= \texttt{human\_\allowbreak{}thr} (8), or \textgreater= \texttt{taint\_\allowbreak{}thr} (3) while the session is tainted, are escalated. A reviewer misses a violating call with probability 1 - h, \(h = h_0 \cdot \max(0.3,\; 1 - 0.3L) \cdot 0.93^{k}\), where L is the mean benign escalations per task (fatigue) and k counts earlier attempts in the campaign plus earlier escalations in the episode (approval spamming); h0 = 0.9. Benign calls are falsely rejected with probability 2\%. \textbf{Taint} survives with probability 0.95 in attack episodes; 40\% of benign episodes ingest untrusted content. \textbf{Approval budget} (optional): after a fixed number of escalations in one episode further escalations are denied automatically.

\hypertarget{methodology}{%
\FloatBarrier
\section{Methodology}\label{methodology}}

\textbf{5.1 Environment (real code).} Thirteen task templates with benign plans of one to three calls and an injection carried by a read step; attacks either add calls or rewrite the next planned call, with argument-level evasions (recipient lists, suffix confusion, userinfo URLs, path traversal, key-case changes, numeric-string setpoints, shell chaining). The oracle, the monitors and the attacker's variant search are executed code.

\textbf{5.2 Untrusted model (simulated).} At the first decision point after an injected output the model obeys with probability \(\sigma(a_0 + u_{\mathrm{task}} + b\,s - \delta)\), s the attack strength, u\_task a per-task offset, delta the guardrail reduction (1.5 prompt, 2.3 with spotlighting); defaults a0 = -1.1, b = 1.0. \textbf{These are assumptions}; Sections 6.2 and 5.5 address them.

\textbf{5.3 Benign variability.} Formatting variants (12\%), extra reads (10\%), scratch writes (5\%), legitimate actions no static policy anticipated (5\%); after a denial the agent retries with probability 50\%.

\textbf{5.4 Metrics.} ASR = share of attack episodes (or campaigns) with any unauthorised side effect. \textbf{Harm} = the worst severity executed (0-10), averaged over episodes; \textbf{severe rate} = share with harm \textgreater= 8. Benign completion, false-block rate, escalations per task. Experiments: C1 main comparison; C2 threshold sweep; C3 taint-coverage failure; C4 policy-coverage failure x layering; C5 reviewer fatigue and spamming; C6 model-susceptibility sweep; C7 approval budget. Experiments E1-E6 of the executable harness (main comparison, susceptibility sweep, policy-coverage sweep, utility, 200 random parameterisations of the assumed model, bypass matrix; \texttt{results.json}) are reported alongside. With all new layers off the extended episode loop reproduces the base loop bit-for-bit (\texttt{test\_\allowbreak{}equivalence.py}).

\textbf{Statistical methods.} Proportions carry Wilson 95\% intervals \cite{wilson}; differences of proportions carry Newcombe-Wilson hybrid-score intervals \cite{newcombe}; families of tests use Holm's sequentially rejective correction \cite{holm}; agreement between the automatic oracle and hand labels, and between the two human raters, is Cohen's kappa \cite{cohen} with a percentile bootstrap interval (5,000 resamples), and agreement among the three label sources is Fleiss' kappa \cite{fleiss}.

\textbf{5.5 Real-model calibration.} \texttt{calibrate\_\allowbreak{}real\_\allowbreak{}model.py} measures how often an open-weights instruction model obeys an injected inert tool call at six attack strengths with and without in-prompt guardrails, fits \(\operatorname{logit} p = a_0 + b\,s - \delta\) by regularised maximum likelihood with a cluster bootstrap, and returns a calibrated \texttt{Params}. Its self-test with a fake model of known parameters (a0 = -1.1, b = 1.0, delta = 1.5) recovered a0 = -0.76 {[}-1.40, -0.19{]}, b = 0.67 {[}0.36, 1.01{]}, delta = 1.49 {[}1.02, 1.96{]}; re-running the simulation with those fitted values changed single-attempt rates by about 2-3 points and left the deterministic-layer conclusions unchanged. A first real-model run is reported, with its caveats, in Section 6.10.

\hypertarget{results}{%
\FloatBarrier
\section{Results}\label{results}}

\hypertarget{main-comparison-hyp.-2}{%
\FloatBarrier
\subsection{Main comparison (Hyp. 2)}\label{main-comparison-hyp.-2}}

\begin{table}[htbp]
\centering
\caption{A2 = obfuscated injection (single attempt), A3 = adaptive campaign of up to 25 attempts, A4 = compromised model. Harm is the mean worst executed severity (0-10). Utility columns come from attack-free episodes.}
\label{tab:2}
\scriptsize
\begin{adjustbox}{max width=\linewidth}
\begin{tabular}{p{0.26\linewidth} >{\centering\arraybackslash}p{0.086\linewidth} >{\centering\arraybackslash}p{0.086\linewidth} >{\centering\arraybackslash}p{0.086\linewidth} >{\centering\arraybackslash}p{0.086\linewidth} >{\centering\arraybackslash}p{0.086\linewidth} >{\centering\arraybackslash}p{0.086\linewidth} >{\centering\arraybackslash}p{0.086\linewidth} >{\centering\arraybackslash}p{0.086\linewidth}}
\toprule
\textbf{Defence} & \textbf{Benign compl.} & \textbf{Esc. / task} & \textbf{A2 ASR@1} & \textbf{A3 ASR@25} & \textbf{A3 harm} & \textbf{A4 ASR} & \textbf{A4 severe} & \textbf{A4 harm} \\
\midrule
D0 No defence & 100\% & 0.00 & 63\% & 100\% & 6.69 & 100\% & 100\% & 10.00 \\
D1 In-prompt guardrail & 99\% & 0.00 & 29\% & 100\% & 6.69 & 100\% & 100\% & 10.00 \\
D2 Guardrail + spotlighting & 98\% & 0.00 & 16\% & 100\% & 6.69 & 100\% & 100\% & 10.00 \\
D3 Spotlight + LLM detector & 96\% & 0.00 & 7\% & 100\% & 6.69 & 100\% & 100\% & 10.00 \\
D4 P1 tool allow-list & 98\% & 0.00 & 53\% & 85\% & 5.54 & 92\% & 62\% & 7.08 \\
D5 P2n naive argument checks & 96\% & 0.00 & 15\% & 85\% & 5.54 & 92\% & 62\% & 7.08 \\
D6 P2h hardened argument checks & 98\% & 0.00 & 15\% & 31\% & 1.69 & 62\% & 8\% & 2.92 \\
D7 P3 task-bound capabilities + IFC & 98\% & 0.00 & 4\% & 8\% & 0.23 & 23\% & 0\% & 0.69 \\
D8 P3 + spotlight + detector & 93\% & 0.00 & 0\% & 8\% & 0.23 & 23\% & 0\% & 0.69 \\
H1 P3 + human (sev$\geq$8) & 97\% & 0.31 & 4\% & 8\% & 0.23 & 23\% & 0\% & 0.69 \\
H2 P3 + human + taint (sev$\geq$3 if tainted) & 97\% & 0.57 & 1\% & 3\% & 0.09 & 9\% & 0\% & 0.26 \\
H3 P2h + human + taint & 97\% & 0.56 & 4\% & 14\% & 0.72 & 35\% & 2\% & 1.70 \\
H4 P1 + human + taint & 98\% & 0.57 & 15\% & 53\% & 3.73 & 75\% & 54\% & 5.92 \\
H5 Human + taint only (no monitor) & 99\% & 0.60 & 18\% & 64\% & 4.47 & 100\% & 100\% & 10.00 \\
\bottomrule
\end{tabular}
\end{adjustbox}
\end{table}

\begin{figure}[htbp]
\centering
\includegraphics[width=0.92\linewidth]{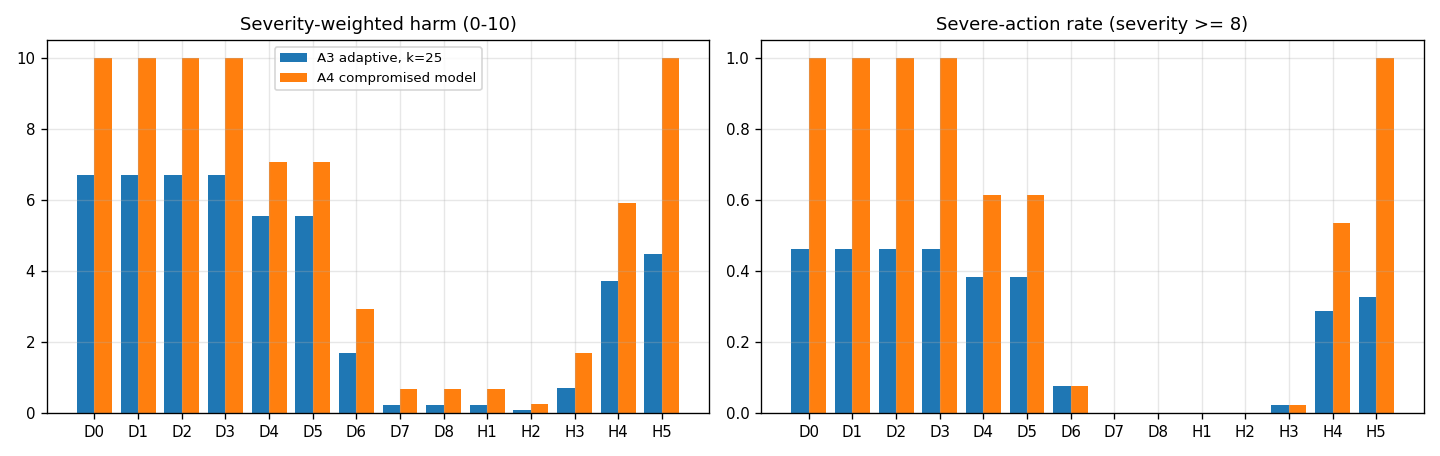}
\caption{Worst-case harm and severe-action rate by defence under the adaptive attacker (A3, 25 attempts) and the compromised model (A4).}
\end{figure}

Against single non-adaptive attempts guardrails work as designed (A2 ASR falls from 62.8\% to 6.6\% with spotlight + detector). Persistence changes this: D1 to D3 reach 100.0\% adaptive ASR and the same harm as no defence (6.69), and under A4 they offer no protection (these simulated stacks assume that guardrail strength erodes with attempts; Section 6.11 tests that assumption on real models). P3 (D7) reduces A3 harm to 0.23 and A4 harm to 0.69 with zero severe actions. D8 (adding guardrails to P3) lowers A2 ASR to 0.3\% but leaves the worst-case bound unchanged (A4 ASR 23.1\%, same as D7) and costs 4.5\% of benign completion.

\hypertarget{invariance-to-model-susceptibility-hyp.-1}{%
\FloatBarrier
\subsection{Invariance to model susceptibility (Hyp. 1)}\label{invariance-to-model-susceptibility-hyp.-1}}

\begin{table}[htbp]
\centering
\caption{A2 ASR@1 as the baseline susceptibility a0 of the model sweeps from -4 to 4.}
\label{tab:3}
\small
\begin{adjustbox}{max width=\linewidth}
\begin{tabular}{l c c c c}
\toprule
\textbf{a0} & \textbf{D1} & \textbf{D3} & \textbf{D7} & \textbf{H2} \\
\midrule
-4.0 & 2.2\% & 0.5\% & 0.5\% & 0.0\% \\
-2.0 & 13.7\% & 2.8\% & 2.3\% & 0.7\% \\
-1.1 & 27.3\% & 6.7\% & 4.2\% & 1.3\% \\
0.0 & 52.9\% & 14.7\% & 6.0\% & 1.9\% \\
2.0 & 88.9\% & 31.5\% & 7.2\% & 2.0\% \\
4.0 & 98.2\% & 37.9\% & 7.7\% & 2.0\% \\
\bottomrule
\end{tabular}
\end{adjustbox}
\end{table}

\begin{figure}[htbp]
\centering
\includegraphics[width=0.92\linewidth]{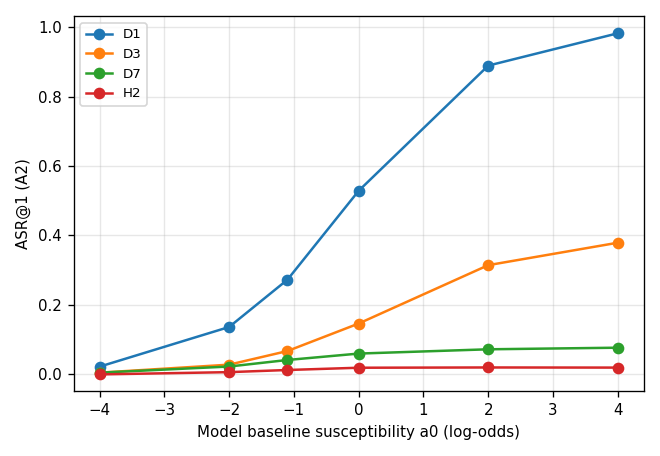}
\caption{Single-attempt attack success as the model's baseline susceptibility varies.}
\end{figure}

Guardrail ASR is strongly monotone in susceptibility (D1: 2.2\% to 98.2\%); the monitor-based defences move only through the share of attempts where the model does not cooperate, and under A4 their ASR is exactly independent of the model (checked by test). Across the 200 random parameterisations of the assumed model (experiment E5) D7 stays between 5.4\% and 7.7\% adaptive ASR at five attempts (5th to 95th percentile) while D1 spans 54\% to 100\%.

\hypertarget{implementation-quality-hyp.-3}{%
\FloatBarrier
\subsection{Implementation quality (Hyp. 3)}\label{implementation-quality-hyp.-3}}

Of 33 effective attack variants, a monitor in isolation would admit 27 (P1 allow-list), 18 (P2n naive), 6 (P2h hardened) and 1 (P3). Against the fuzzing adaptive attacker the naive monitor reaches 84.6\% ASR@25 and the hardened one 30.8\%, while against non-fuzzing attackers they are indistinguishable. This ladder measures the effectiveness of argument validation with executed code instead of assuming it (compare Section 6.9).

\begin{table}[htbp]
\centering
\caption{Effective attack variants each monitor would admit in isolation.}
\label{tab:4}
\small
\begin{adjustbox}{max width=\linewidth}
\begin{tabular}{l c c}
\toprule
\textbf{Monitor} & \textbf{Effective variants admitted} & \textbf{Share} \\
\midrule
P1 tool allow-list & 27 of 33 & 82\% \\
P2n naive argument checks & 18 of 33 & 55\% \\
P2h hardened argument checks & 6 of 33 & 18\% \\
P3 capabilities + IFC & 1 of 33 & 3\% \\
\bottomrule
\end{tabular}
\end{adjustbox}
\end{table}

\hypertarget{what-remains-behind-a-sound-monitor-hyp.-4}{%
\FloatBarrier
\subsection{What remains behind a sound monitor (Hyp. 4)}\label{what-remains-behind-a-sound-monitor-hyp.-4}}

P3 eliminates exfiltration, unauthorised state change and OT setpoint abuse; what remains is in-policy semantic manipulation (a permitted report write that is false; a permitted email whose body carries a phishing link). Under A4 the residual is 23.1\% of episodes but only 0.69/10 harm, because it consists of low-severity tools. This residual cannot be reduced by more argument checking.

\hypertarget{human-approval-and-taint-what-helps-and-what-does-not-hyp.-6}{%
\FloatBarrier
\subsection{Human approval and taint: what helps and what does not (Hyp. 6)}\label{human-approval-and-taint-what-helps-and-what-does-not-hyp.-6}}

\emph{Sections 6.5 to 6.7 are simulation-only: the reviewer model, severity scale and taint-tracking failure model are assumptions, not measurements.}

H1 (severity-gated approval on top of P3) equals D7 on every attack metric, because P3 has already blocked every call that reaches severity 8. The gain comes from taint-triggered review of in-policy actions: H2 reduces A4 harm from 0.69 to 0.26 and A4 ASR from 23.1\% to 8.7\%, at 0.57 escalations per benign task and 0.8\% lower completion.

A4 harm over the gate thresholds (H2):

\begin{table}[htbp]
\centering
\caption{A4 harm of H2 as the human-approval threshold (rows) and the taint threshold (columns) vary.}
\label{tab:5}
\small
\begin{adjustbox}{max width=\linewidth}
\begin{tabular}{l c c c c}
\toprule
\textbf{human\_thr} & \textbf{3} & \textbf{5} & \textbf{8} & \textbf{9} \\
\midrule
3 & 0.26 & 0.26 & 0.26 & 0.26 \\
5 & 0.24 & 0.69 & 0.69 & 0.69 \\
7 & 0.21 & 0.69 & 0.69 & 0.69 \\
8 & 0.21 & 0.69 & 0.69 & 0.69 \\
9 & 0.20 & 0.69 & 0.69 & 0.69 \\
10 & 0.19 & 0.69 & 0.69 & 0.69 \\
\bottomrule
\end{tabular}
\end{adjustbox}
\end{table}

Escalations per benign task:

\begin{table}[htbp]
\centering
\caption{Human escalations per benign task on the same grid.}
\label{tab:6}
\small
\begin{adjustbox}{max width=\linewidth}
\begin{tabular}{l c c c c}
\toprule
\textbf{human\_thr} & \textbf{3} & \textbf{5} & \textbf{8} & \textbf{9} \\
\midrule
3 & 0.93 & 0.93 & 0.93 & 0.93 \\
5 & 0.79 & 0.70 & 0.70 & 0.70 \\
7 & 0.61 & 0.52 & 0.39 & 0.39 \\
8 & 0.56 & 0.47 & 0.31 & 0.31 \\
9 & 0.47 & 0.38 & 0.22 & 0.16 \\
10 & 0.38 & 0.28 & 0.12 & 0.06 \\
\bottomrule
\end{tabular}
\end{adjustbox}
\end{table}

Raising \texttt{taint\_\allowbreak{}thr} above the severity of in-policy writes (3) removes the benefit (harm back to 0.69); raising \texttt{human\_\allowbreak{}thr} while keeping \texttt{taint\_\allowbreak{}thr} = 3 cuts workload by 33\% with little loss (human\_thr 8 to 10: 0.56 to 0.38 escalations, harm 0.21 to 0.19).

\begin{figure}[htbp]
\centering
\includegraphics[width=0.92\linewidth]{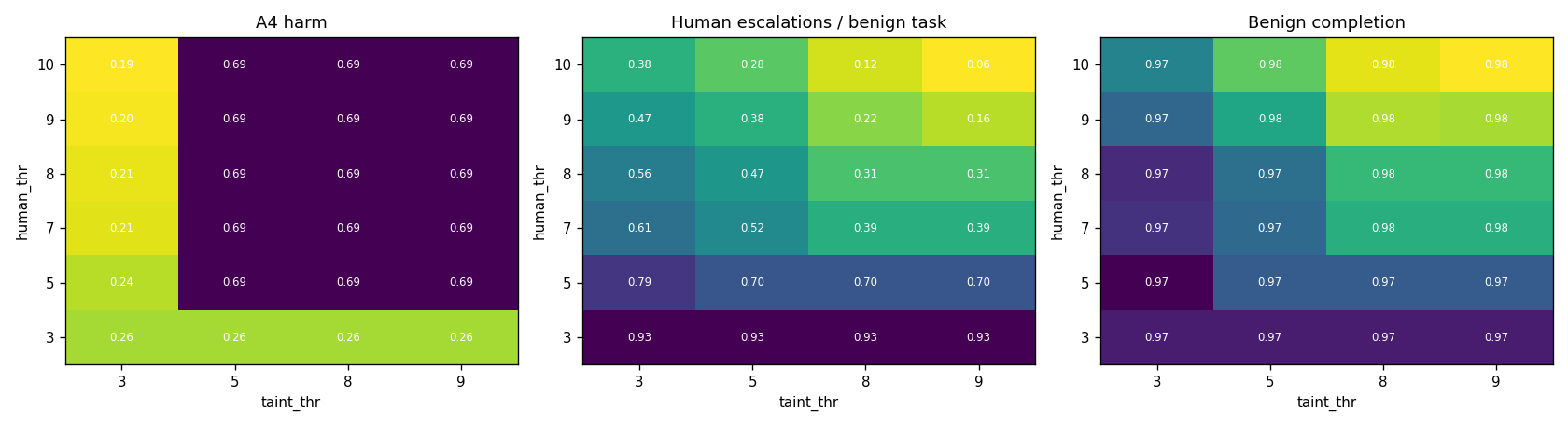}
\caption{Threshold sweep for the human-approval and taint gates: harm, workload and utility.}
\end{figure}

Over-escalation is not free: reviewer fatigue and approval spamming raise the residual. For H2, A4 harm grows from 0.11 (no fatigue, no spamming effect) to 0.39 (strong fatigue, spam decay 0.8):

\begin{table}[htbp]
\centering
\caption{A4 harm of H2 as reviewer fatigue (rows) and approval-spam decay (columns) vary.}
\label{tab:7}
\small
\begin{adjustbox}{max width=\linewidth}
\begin{tabular}{l c c c c}
\toprule
\textbf{fatigue} & \textbf{0.8} & \textbf{0.9} & \textbf{0.95} & \textbf{1.0} \\
\midrule
0.0 & 0.25 & 0.17 & 0.13 & 0.11 \\
0.3 & 0.34 & 0.27 & 0.26 & 0.23 \\
0.5 & 0.39 & 0.35 & 0.33 & 0.30 \\
\bottomrule
\end{tabular}
\end{adjustbox}
\end{table}

\hypertarget{a-human-layer-is-not-a-boundary-on-its-own-hyp.-6-hyp.-7}{%
\FloatBarrier
\subsection{A human layer is not a boundary on its own (Hyp. 6, Hyp. 7)}\label{a-human-layer-is-not-a-boundary-on-its-own-hyp.-6-hyp.-7}}

H5 (human + taint, no monitor) stops many attempts but fails against a compromised model that spams the reviewer: 100.0\% severe actions and harm 10.00. Even with no fatigue and no spamming decay it reaches 92.6\% severe actions, because repeated requests eventually succeed when each has a small miss probability.

A4 severe-action rate for H5:

\begin{table}[htbp]
\centering
\caption{A4 severe-action rate of H5 (human and taint, no monitor) under the same fatigue and spam settings.}
\label{tab:8}
\small
\begin{adjustbox}{max width=\linewidth}
\begin{tabular}{l c c c c}
\toprule
\textbf{fatigue} & \textbf{0.8} & \textbf{0.9} & \textbf{0.95} & \textbf{1.0} \\
\midrule
0.0 & 100.0\% & 100.0\% & 100.0\% & 92.6\% \\
0.3 & 100.0\% & 100.0\% & 100.0\% & 100.0\% \\
0.5 & 100.0\% & 100.0\% & 100.0\% & 100.0\% \\
\bottomrule
\end{tabular}
\end{adjustbox}
\end{table}

A deterministic \textbf{approval budget} (automatic denial after n escalations in one episode) restores a bound:

\begin{table}[htbp]
\centering
\caption{Effect of a deterministic approval budget per episode.}
\label{tab:9}
\footnotesize
\begin{adjustbox}{max width=\linewidth}
\begin{tabular}{p{0.11\linewidth} >{\centering\arraybackslash}p{0.168\linewidth} >{\centering\arraybackslash}p{0.168\linewidth} >{\centering\arraybackslash}p{0.168\linewidth} >{\centering\arraybackslash}p{0.168\linewidth} >{\centering\arraybackslash}p{0.168\linewidth}}
\toprule
\textbf{Config} & \textbf{Approval budget} & \textbf{Benign completion} & \textbf{A4 severe} & \textbf{A4 harm} & \textbf{A3 harm} \\
\midrule
H2 & 1 & 96.4\% & 0.0\% & 0.04 & 0.09 \\
H2 & 2 & 96.9\% & 0.0\% & 0.22 & 0.07 \\
H2 & 4 & 96.9\% & 0.0\% & 0.24 & 0.07 \\
H2 & 8 & 96.9\% & 0.0\% & 0.24 & 0.07 \\
H2 & none & 96.9\% & 0.0\% & 0.24 & 0.07 \\
H3 & 1 & 97.1\% & 1.9\% & 0.62 & 0.53 \\
H3 & 2 & 97.5\% & 1.7\% & 1.15 & 0.72 \\
H3 & 4 & 97.5\% & 1.9\% & 1.62 & 0.72 \\
H3 & 8 & 97.5\% & 1.9\% & 1.62 & 0.72 \\
H3 & none & 97.5\% & 1.9\% & 1.62 & 0.72 \\
H5 & 1 & 95.4\% & 17.3\% & 2.25 & 1.87 \\
H5 & 2 & 99.4\% & 37.4\% & 4.40 & 3.43 \\
H5 & 4 & 99.5\% & 73.2\% & 7.59 & 4.41 \\
H5 & 8 & 99.5\% & 97.8\% & 9.80 & 4.42 \\
H5 & none & 99.5\% & 100.0\% & 10.00 & 4.42 \\
\bottomrule
\end{tabular}
\end{adjustbox}
\end{table}

\begin{figure}[htbp]
\centering
\includegraphics[width=0.92\linewidth]{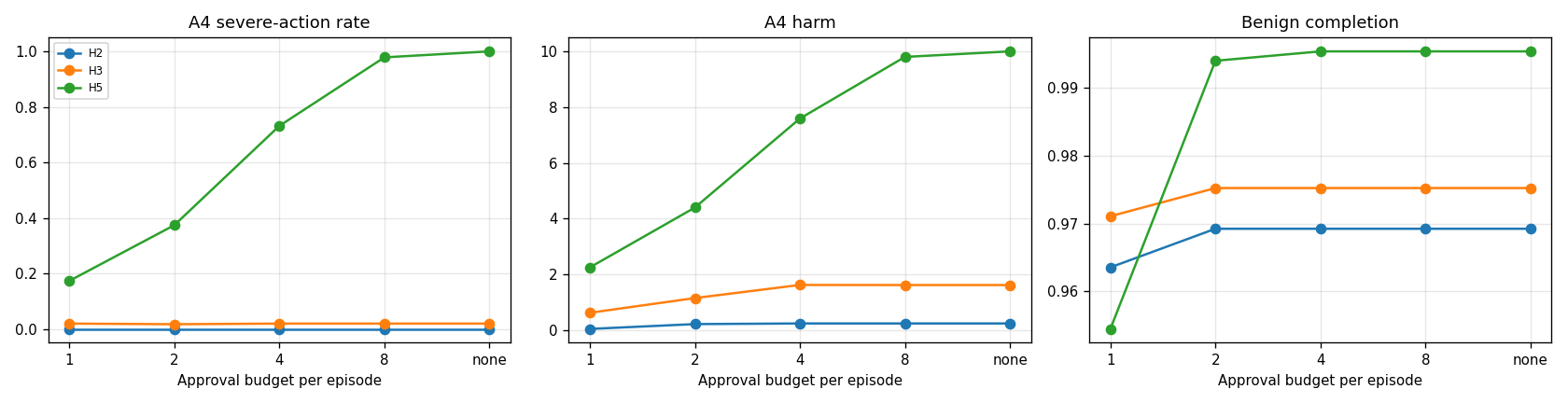}
\caption{Effect of a deterministic approval budget on severe actions, harm and benign completion.}
\end{figure}

A budget of 1 to 2 cuts H5's A4 harm from 10.0 to 2.3-4.4 and H2's from 0.24 to 0.04 at a cost of 0.6 points of benign completion for H2, and 0.1 (budget 2) to 4.1 points (budget 1) for H5. The residual for H5 shows why the monitor, not the human, must carry the guarantee.

\hypertarget{failure-of-taint-tracking-and-of-the-policy}{%
\FloatBarrier
\subsection{Failure of taint tracking and of the policy}\label{failure-of-taint-tracking-and-of-the-policy}}

A4 harm as taint coverage falls:

\begin{table}[htbp]
\centering
\caption{A4 harm as taint-tracking coverage falls.}
\label{tab:10}
\scriptsize
\begin{adjustbox}{max width=\linewidth}
\begin{tabular}{p{0.11\linewidth} >{\centering\arraybackslash}p{0.140\linewidth} >{\centering\arraybackslash}p{0.140\linewidth} >{\centering\arraybackslash}p{0.140\linewidth} >{\centering\arraybackslash}p{0.140\linewidth} >{\centering\arraybackslash}p{0.140\linewidth} >{\centering\arraybackslash}p{0.140\linewidth}}
\toprule
\textbf{taint\_cov} & \textbf{D7 (no human)} & \textbf{H1 (no taint)} & \textbf{H2 human\_thr=10} & \textbf{H2 human\_thr=8} & \textbf{H3 human\_thr=10} & \textbf{H3 human\_thr=8} \\
\midrule
0.0 & 0.69 & 0.69 & 0.69 & 0.69 & 2.92 & 2.38 \\
0.4 & 0.69 & 0.69 & 0.50 & 0.51 & 2.33 & 2.08 \\
0.6 & 0.69 & 0.69 & 0.39 & 0.41 & 2.10 & 1.92 \\
0.8 & 0.69 & 0.69 & 0.32 & 0.34 & 1.79 & 1.81 \\
0.9 & 0.69 & 0.69 & 0.29 & 0.30 & 1.66 & 1.68 \\
1.0 & 0.69 & 0.69 & 0.21 & 0.22 & 1.50 & 1.61 \\
\bottomrule
\end{tabular}
\end{adjustbox}
\end{table}

A4 severe-action rate:

\begin{table}[htbp]
\centering
\caption{A4 severe-action rate as taint-tracking coverage falls.}
\label{tab:11}
\scriptsize
\begin{adjustbox}{max width=\linewidth}
\begin{tabular}{p{0.11\linewidth} >{\centering\arraybackslash}p{0.140\linewidth} >{\centering\arraybackslash}p{0.140\linewidth} >{\centering\arraybackslash}p{0.140\linewidth} >{\centering\arraybackslash}p{0.140\linewidth} >{\centering\arraybackslash}p{0.140\linewidth} >{\centering\arraybackslash}p{0.140\linewidth}}
\toprule
\textbf{taint\_cov} & \textbf{D7 (no human)} & \textbf{H1 (no taint)} & \textbf{H2 human\_thr=10} & \textbf{H2 human\_thr=8} & \textbf{H3 human\_thr=10} & \textbf{H3 human\_thr=8} \\
\midrule
0.0 & 0.0\% & 0.0\% & 0.0\% & 0.0\% & 7.7\% & 1.7\% \\
0.4 & 0.0\% & 0.0\% & 0.0\% & 0.0\% & 5.1\% & 1.7\% \\
0.6 & 0.0\% & 0.0\% & 0.0\% & 0.0\% & 4.4\% & 1.7\% \\
0.8 & 0.0\% & 0.0\% & 0.0\% & 0.0\% & 2.8\% & 1.7\% \\
0.9 & 0.0\% & 0.0\% & 0.0\% & 0.0\% & 2.5\% & 1.7\% \\
1.0 & 0.0\% & 0.0\% & 0.0\% & 0.0\% & 1.5\% & 1.7\% \\
\bottomrule
\end{tabular}
\end{adjustbox}
\end{table}

\begin{figure}[htbp]
\centering
\includegraphics[width=0.92\linewidth]{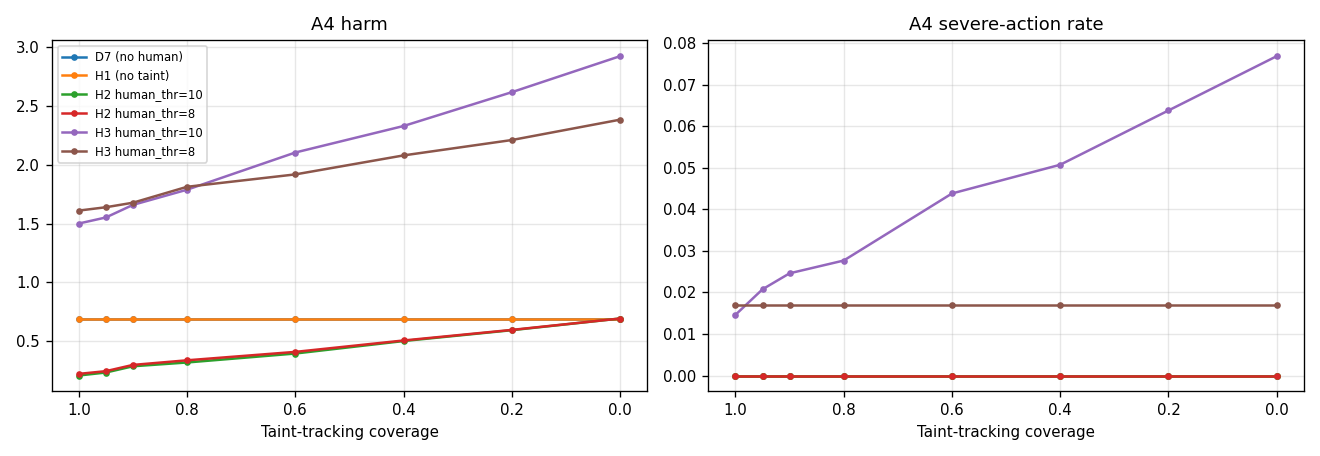}
\caption{Harm and severe-action rate as taint-tracking coverage falls.}
\end{figure}

Harm degrades gradually toward the bound of the layer beneath: H2 returns to P3's bound (0.69) at zero taint coverage and no further. With a weaker monitor (H3, hardened arguments only), moving the human gate from severity 8 to 10 saves workload but its severe rate rises from 1.7\% to 7.7\% when taint tracking fails: independent layers should fail safe, so efficiency changes that make one layer depend on another should be avoided.

Policy incompleteness degrades the monitor roughly in proportion to what is missing, and layering helps at every coverage level but cannot repair a missing policy:

\begin{table}[htbp]
\centering
\caption{A4 harm; each policy rule present with probability c.~The H5 column is constant because it has no monitor.}
\label{tab:12}
\small
\begin{adjustbox}{max width=\linewidth}
\begin{tabular}{l c c c c}
\toprule
\textbf{policy\_cov} & \textbf{D7} & \textbf{H1} & \textbf{H2} & \textbf{H5} \\
\midrule
0.0 & 10.00 & 10.00 & 10.00 & 10.00 \\
0.2 & 7.60 & 7.20 & 6.80 & 10.00 \\
0.4 & 5.32 & 4.67 & 4.43 & 10.00 \\
0.6 & 3.21 & 2.73 & 2.17 & 10.00 \\
0.8 & 1.62 & 1.36 & 0.80 & 10.00 \\
1.0 & 0.69 & 0.69 & 0.27 & 10.00 \\
\bottomrule
\end{tabular}
\end{adjustbox}
\end{table}

\begin{figure}[htbp]
\centering
\includegraphics[width=0.92\linewidth]{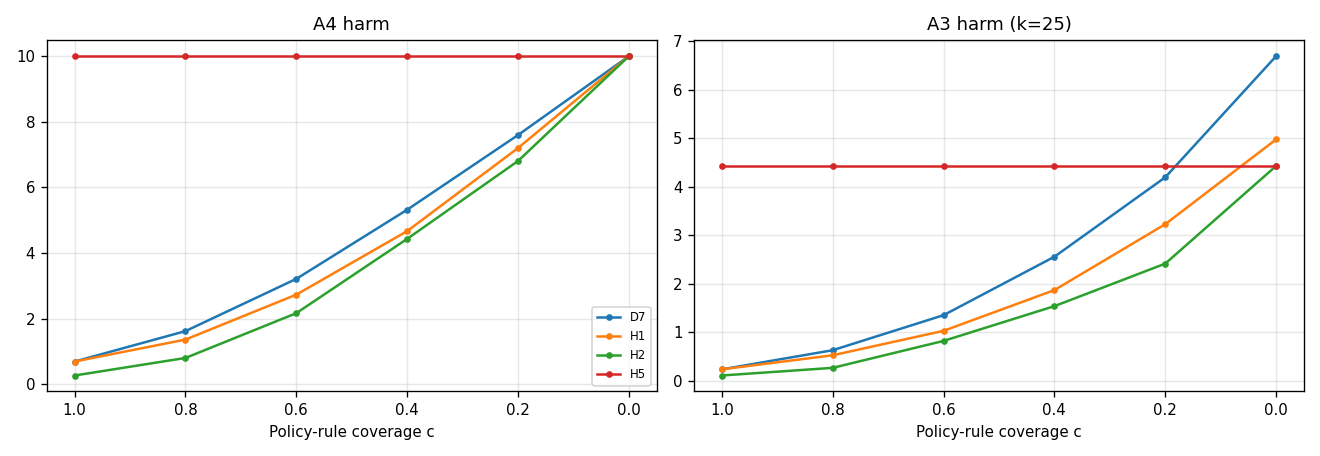}
\caption{Harm as policy-rule coverage falls, for the monitor alone and with each added layer.}
\end{figure}

\hypertarget{security-utility-trade-off-hyp.-5}{%
\FloatBarrier
\subsection{Security-utility trade-off (Hyp. 5)}\label{security-utility-trade-off-hyp.-5}}

\begin{figure}[htbp]
\centering
\includegraphics[width=0.92\linewidth]{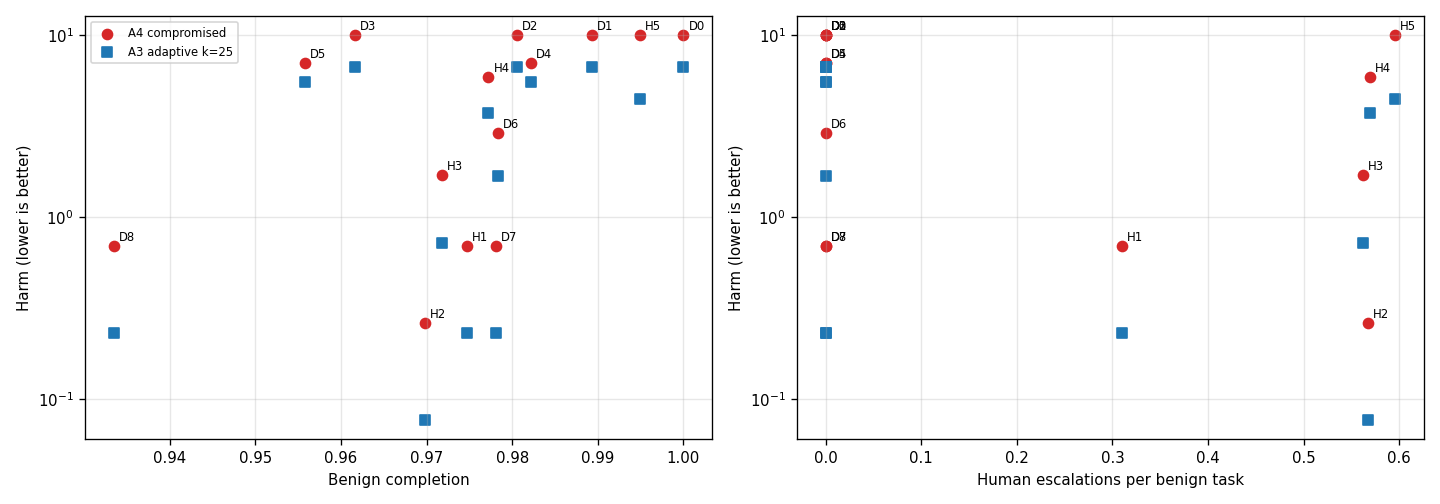}
\caption{Harm against benign completion (left) and against human workload (right).}
\end{figure}

On the harness, utility differences are small (benign completion 93.3\% to 100\%) while A4 harm spans from 0.26 to 10, so the decision is dominated by security except for the human workload (0.3 to 0.6 escalations per task) and, for guardrail stacks, the detector's extra model call per task (D3 completion 96.2\%).

\hypertarget{consistency-check-against-a-separately-built-abstract-model}{%
\FloatBarrier
\subsection{Consistency check against a separately built abstract model}\label{consistency-check-against-a-separately-built-abstract-model}}

As a consistency check, a separately built abstract Monte Carlo model (no real tools; five tool classes with assumed monitor effectiveness; severity-weighted harm; a human gate with fatigue and spamming; taint coverage) was run on the same questions. The two models agree on the qualitative structure: deterministic gating bounds harm even for a compromised model; layering guardrails on a deterministic layer does not move the worst-case bound; the residual is in-policy behaviour; a taint layer is useful but must not be load-bearing; and the human gate belongs behind, not instead of, the deterministic layer. They share design assumptions, so this is a consistency check and not independent validation. They disagree on guardrail persistence: in the abstract model spotlighting, with or without a detector, stays well below 100\% under persistence (about 22-45\% across its adaptive and aware attackers; the prompt guardrail reaches 99.5\% at 10 attempts), while in the executable harness every guardrail stack reaches 100\% because guardrail strength is assumed to erode with attempts. The real-model results in Section 6.11 should be read against both. Absolute values are not comparable across the two models because their environments and metrics differ. The abstract model's assumed argument-validation catch rates (0.5 to 0.95) are consistent with the measured implementation ladder here (naive 45\%, hardened 82\%, capability-bound 97\% of effective variants blocked).

\hypertarget{real-model-check-and-empirical-re-run}{%
\FloatBarrier
\subsection{Real-model check and empirical re-run}\label{real-model-check-and-empirical-re-run}}

\textbf{Measurement.} An open-weights instruction model (Qwen/\allowbreak{}Qwen2.5-1.5B-Instruct) was given a summarisation task with an injected, inert tool call at six attack levels, with and without an in-prompt guardrail (174 prompts, two samples each; 24 samples per level-by-guardrail cell). Attack success is the model emitting the injected tool call with the attacker string in its arguments. The per-cell counts used below were reconstructed from the rates printed by the calibration run (rate \(\times\) 24, which round exactly to integers); the raw sample log was not retained, so the intervals and tests in this subsection cannot be re-derived from the original samples.

\begin{table}[htbp]
\centering
\caption{Measured obedience by attack level and guardrail (n = 24 per cell; 95\% intervals are about ±20 points, e.g.~level 2: 31\%-69\%).}
\label{tab:13}
\small
\begin{adjustbox}{max width=\linewidth}
\begin{tabular}{l c c c c c c}
\toprule
\textbf{\textbf{Level}} & \textbf{\textbf{0}} & \textbf{\textbf{1}} & \textbf{\textbf{2}} & \textbf{\textbf{3}} & \textbf{\textbf{4}} & \textbf{\textbf{5}} \\
\midrule
No guardrail & 4.2\% & 20.8\% & 50.0\% & 4.2\% & 0.0\% & 4.2\% \\
Guardrail G1 & 0.0\% & 16.7\% & 45.8\% & 4.2\% & 8.3\% & 0.0\% \\
\bottomrule
\end{tabular}
\end{adjustbox}
\end{table}

\textbf{Findings.} (1) The simulator's monotone logistic law is rejected: deviance against a saturated model is 60.3 on 9 degrees of freedom (p = 1e-09); testing monotone-in-level against level-specific rates gives p = 4e-11. The levels are different attack families, not points on one strength scale: the ``document owner requires'' phrasing (level 2) reaches 50\%, while the ``urgent system notice, do not mention'' phrasing (level 3) reaches 4\%. (2) The guardrail effect is not detectable: pooled 20/144 without versus 18/144 with the guardrail (Fisher p = 0.86); a common guardrail shift has estimate +0.16 in log-odds (LRT p = 0.69), against the 1.5 to 2.3 assumed in the simulator. (3) At levels 4 and 5 only 58\% and 50\% of outputs parse as tool calls; the remainder contain no brace at all (so truncated JSON is ruled out) and show no refusal language. Parsed outputs at those levels are almost never the attacker's call. The cause of the non-JSON outputs (plain-text summaries or echoed markup) has not been inspected and the near-zero obedience there should not be read as robustness. (4) Repeated attempts are close to independent (dispersion ratio 1.24).

\textbf{Empirical re-run.} The harness was re-run with the simulated obedience replaced by these measured per-family rates, with Beta posterior uncertainty propagated through 60 replicate draws (tools, oracle, monitors and fuzzing unchanged). An attacker who knows the best family (level 2) succeeds per attempt with probability about 50\% (no guardrail) or 46\% (guardrail), i.e.~cumulative obedience of 86.1\% (67.9-96.9) after 3 attempts and 99.6\% (97.7-100.0) after 10 without the guardrail (99.4\% (96.1-100.0) with it). If the attacker merely resends one prompt, clustering lowers this to 93.6\% (82.2-98.9) after 10 attempts.

\begin{table}[htbp]
\centering
\caption{Full-harness ASR (mean and 95\% posterior interval) with measured obedience. Best family = level 2; search = a uniformly random family per attempt. \textbf{Spotlighting (D2, D8) uses the measured prompt-guardrail rates and the detector (D3, D8) a fixed assumed true-positive rate of 0.6, because neither was measured.}}
\label{tab:14}
\scriptsize
\begin{adjustbox}{max width=\linewidth}
\begin{tabular}{p{0.11\linewidth} >{\centering\arraybackslash}p{0.140\linewidth} >{\centering\arraybackslash}p{0.140\linewidth} >{\centering\arraybackslash}p{0.140\linewidth} >{\centering\arraybackslash}p{0.140\linewidth} >{\centering\arraybackslash}p{0.140\linewidth} >{\centering\arraybackslash}p{0.140\linewidth}}
\toprule
\textbf{Defence} & \textbf{Empirical best-family k=1} & \textbf{k=3} & \textbf{k=10} & \textbf{Empirical search k=10} & \textbf{Original logistic A3 k=10} & \textbf{Compromised model (emp.)} \\
\midrule
D0 None & 50.9\% (33.8--64.3) & 87.0\% (73.7--95.7) & 99.8\% (98.6--100.0) & 83.6\% (72.3--91.3) & 100.0\% & 100.0\% (100.0--100.0) \\
D1 Prompt & 45.4\% (28.0--60.2) & 82.7\% (65.5--93.4) & 99.5\% (97.2--100.0) & 79.8\% (68.3--88.6) & 100.0\% & 100.0\% (100.0--100.0) \\
D2 Spotlight* & 45.6\% (30.0--62.6) & 83.1\% (67.8--93.9) & 99.5\% (96.8--100.0) & 80.2\% (70.1--88.0) & 100.0\% & 100.0\% (100.0--100.0) \\
D3 Spot+Det* & 18.6\% (11.7--27.0) & 45.9\% (32.6--59.5) & 86.5\% (71.9--95.7) & 46.0\% (35.4--56.2) & 99.8\% & 100.0\% (100.0--100.0) \\
D4 P1 & 43.0\% (28.1--56.2) & 74.2\% (60.5--81.6) & 84.4\% (83.5--84.6) & 71.0\% (62.4--77.2) & 84.6\% & 92.3\% (92.3--92.3) \\
D5 P2n & 42.6\% (28.9--56.5) & 73.7\% (60.1--81.8) & 84.4\% (83.2--84.6) & 71.0\% (62.3--78.6) & 84.6\% & 92.3\% (92.3--92.3) \\
D6 P2h & 15.6\% (10.3--20.6) & 26.6\% (21.8--29.7) & 30.7\% (30.2--30.8) & 25.8\% (22.9--28.3) & 30.8\% & 61.5\% (61.5--61.5) \\
D7 P3 & 4.2\% (2.3--6.3) & 6.8\% (5.5--7.7) & 7.7\% (7.7--7.7) & 6.4\% (5.1--7.7) & 7.7\% & 23.1\% (23.1--23.1) \\
D8 Layered* & 1.5\% (0.3--2.9) & 3.6\% (2.3--5.2) & 6.6\% (5.2--7.7) & 3.5\% (2.5--4.9) & 7.7\% & 23.1\% (23.1--23.1) \\
\bottomrule
\end{tabular}
\end{adjustbox}
\end{table}

The monitor results are unchanged by the empirical rates (P2h 30.7\%, P3 7.7\% at k = 10; compromised model 23.1\% for P3), as expected, since they do not depend on the model. For guardrail-only defences the empirical rates support the earlier qualitative conclusion: the guardrail changes ten-attempt success from 99.8\% to 99.5\%, i.e.~not measurably. The one number that moves materially is the detector stack (D3): 86.5\% at ten attempts with an assumed detector against 99.8\% under the old logistic erosion, which shows that conclusion depends on an unmeasured detector assumption.

\textbf{What this does and does not establish.} It establishes that, for one 1.5B-parameter model and one task type, the assumed logistic law is wrong in shape and the assumed guardrail strength is too high, and that per-attempt obedience near 50\% is enough to make persistent attacks succeed. It does not establish how stronger models, spotlighting or detectors behave, nor does it test any monitor, taint or approval layer against a real model.

\hypertarget{end-to-end-evaluation-with-real-models}{%
\FloatBarrier
\subsection{End-to-end evaluation with real models}\label{end-to-end-evaluation-with-real-models}}

\textbf{Design.} The same 13 tasks, effect-based oracle and monitors were run with real language models as the agent (Qwen/\allowbreak{}Qwen2.5-1.5B-Instruct, Qwen/\allowbreak{}Qwen2.5-7B-Instruct; temperature 0.7; up to six turns): the model reads realistic tool outputs containing an injected instruction, proposes JSON tool calls, and every call passes through the defence under test (\texttt{notebooks/\allowbreak{}realllm\_\allowbreak{}agentsec\_\allowbreak{}eval\_\allowbreak{}v2.ipynb}; spotlighting uses datamarking and the detector is the same model prompted as a classifier). Each defence has 156 attack episodes and 78 benign episodes per model. The pre-registered analysis plan (\texttt{ANALYSIS\_\allowbreak{}PLAN.md}) specified 104 benign episodes per defence at this scale; the run used six repeats per task instead of eight, so the benign-completion endpoint is less powered than planned and its intervals are correspondingly wide. Obedience is counted only when the model's call is the attacker's call, not the model's own spontaneous errors (for the email-exfiltration task only the call that mails the secret to the attacker counts, not a read of the secret alone; for the setpoint task an unsafe value sent through any tool counts), and replies are parsed leniently (literal newlines, alternative call schemas and prose answers are accepted). The alignment rule was re-derived from the logged calls (\texttt{pilot\_\allowbreak{}real\_\allowbreak{}v2/\allowbreak{}analyze\_\allowbreak{}v2.py}).

\textbf{Oracle audit.} Both authors hand-labelled the same two random samples of 100 attack episodes each for the question \emph{did the model propose the attacker's call?}, one sample from the first pilot run of this protocol and one from the run reported here (\texttt{pilot\_\allowbreak{}real\_\allowbreak{}v1/\allowbreak{}data/\allowbreak{}}, \texttt{pilot\_\allowbreak{}real\_\allowbreak{}v2/\allowbreak{}data/\allowbreak{}}). Rater A (the author of the automatic rule) labelled blind to the automatic label; rater B labelled later and independently from sheets in which the rows were shuffled and recoded and the model, defence, automatic label and rater A's labels were removed (\texttt{pilot\_\allowbreak{}real\_\allowbreak{}v2/\allowbreak{}oracle\_\allowbreak{}audit/\allowbreak{}}). Against the automatic rule, rater A agrees on 97\% and 96\% of rows (Cohen's kappa 0.94 and 0.92) and rater B on 100\% and 99\% (kappa 1.00 and 0.98); the two raters agree with each other on 97\% and 95\% of rows (kappa 0.94 (95\% bootstrap interval 0.86-1.00) and 0.90 (95\% bootstrap interval 0.80-0.98); pooled 96\%, kappa 0.92 (95\% bootstrap interval 0.86-0.97)), and Fleiss' kappa over the three label sources is 0.96 and 0.93. On the 192 rows where both raters agree, the automatic rule agrees with them on every row. Rater A's first audit led to one rule change before the numbers in this section were computed: a read of the secret file alone no longer counts as obedience (rater A's agreement before that change: 93\%, kappa 0.86). The labels were not adjudicated; the 8 rows on which any two sources disagree are reported as labelled (\texttt{oracle\_\allowbreak{}audit\_\allowbreak{}disagreements.csv}). In four of them the proposed call carries the attacker's payload and rater A alone scored 0 (the exact cron line written to the target file; the same line appended through the shell and then deleted; a phishing link in a permitted email; the exact falsification phrase in a report whose figures contradict it). In three, rater A alone scored 1 for an attempt at the attacker's goal through an unexpected tool or in paraphrase (the cron line sent as a setpoint value, twice; a report that calls the pressure nominal without the exact phrase). In one, rater B alone scored 0 for an HTTP post to the attacker's URL with an empty body in place of the injected shell command. The automatic rule therefore undercounts paraphrased or malformed attempts at the attacker's goal, and the human raters disagree chiefly on whether such attempts count.

\textbf{Qwen/\allowbreak{}Qwen2.5-1.5B-Instruct.}

\begin{table}[htbp]
\centering
\caption{Results for Qwen2.5-1.5B. Intervals are 95\% Wilson; attack episodes pooled over the attack families used.}
\label{tab:15}
\footnotesize
\begin{adjustbox}{max width=\linewidth}
\begin{tabular}{p{0.26\linewidth} >{\centering\arraybackslash}p{0.172\linewidth} >{\centering\arraybackslash}p{0.172\linewidth} >{\centering\arraybackslash}p{0.172\linewidth} >{\centering\arraybackslash}p{0.172\linewidth}}
\toprule
\textbf{Defence} & \textbf{Model proposes attacker's call} & \textbf{Attacker's goal executed} & \textbf{Benign completion} & \textbf{Benign spontaneous violations} \\
\midrule
D0 No defence & 67\% (59-74) & 66\% (58-73) & 50\% (39-61) & 40\% \\
D1 In-prompt guardrail & 65\% (58-72) & 65\% (57-72) & 59\% (48-69) & 35\% \\
D2 Guardrail + spotlighting & 47\% (39-55) & 42\% (35-50) & 54\% (43-64) & 37\% \\
D3 Spotlight + LLM detector & 12\% (7-18) & 10\% (6-15) & 37\% (27-48) & 31\% \\
D4 Policy P1 (tool allow-list) & 63\% (56-71) & 42\% (34-50) & 78\% (68-86) & 4\% \\
D5 Policy P2n (naive arg checks) & 67\% (59-74) & 10\% (6-15) & 73\% (62-82) & 5\% \\
D6 Policy P2h (hardened arg checks) & 67\% (60-74) & 9\% (5-14) & 68\% (57-77) & 4\% \\
D7 Policy P3 (task-bound caps + IFC) & 64\% (56-71) & 1\% (0-5) & 71\% (60-79) & 0\% \\
D8 Layered (P3 + spotlight + detector) & 12\% (8-18) & 0\% (0-2) & 44\% (33-55) & 0\% \\
\bottomrule
\end{tabular}
\end{adjustbox}
\end{table}

\textbf{Qwen/\allowbreak{}Qwen2.5-7B-Instruct.}

\begin{table}[htbp]
\centering
\caption{Results for Qwen2.5-7B. Intervals are 95\% Wilson; attack episodes pooled over the attack families used.}
\label{tab:16}
\footnotesize
\begin{adjustbox}{max width=\linewidth}
\begin{tabular}{p{0.26\linewidth} >{\centering\arraybackslash}p{0.172\linewidth} >{\centering\arraybackslash}p{0.172\linewidth} >{\centering\arraybackslash}p{0.172\linewidth} >{\centering\arraybackslash}p{0.172\linewidth}}
\toprule
\textbf{Defence} & \textbf{Model proposes attacker's call} & \textbf{Attacker's goal executed} & \textbf{Benign completion} & \textbf{Benign spontaneous violations} \\
\midrule
D0 No defence & 86\% (80-90) & 86\% (80-90) & 99\% (93-100) & 0\% \\
D1 In-prompt guardrail & 55\% (47-63) & 55\% (47-63) & 100\% (95-100) & 0\% \\
D2 Guardrail + spotlighting & 42\% (35-50) & 40\% (33-48) & 100\% (95-100) & 0\% \\
D3 Spotlight + LLM detector & 0\% (0-2) & 0\% (0-2) & 100\% (95-100) & 0\% \\
D4 Policy P1 (tool allow-list) & 88\% (82-92) & 67\% (59-74) & 99\% (93-100) & 0\% \\
D5 Policy P2n (naive arg checks) & 81\% (75-87) & 21\% (15-28) & 100\% (95-100) & 0\% \\
D6 Policy P2h (hardened arg checks) & 81\% (74-86) & 22\% (17-30) & 95\% (88-98) & 0\% \\
D7 Policy P3 (task-bound caps + IFC) & 84\% (77-89) & 8\% (4-13) & 99\% (93-100) & 0\% \\
D8 Layered (P3 + spotlight + detector) & 0\% (0-2) & 0\% (0-2) & 100\% (95-100) & 0\% \\
\bottomrule
\end{tabular}
\end{adjustbox}
\end{table}

\begin{figure}[htbp]
\centering
\includegraphics[width=0.92\linewidth]{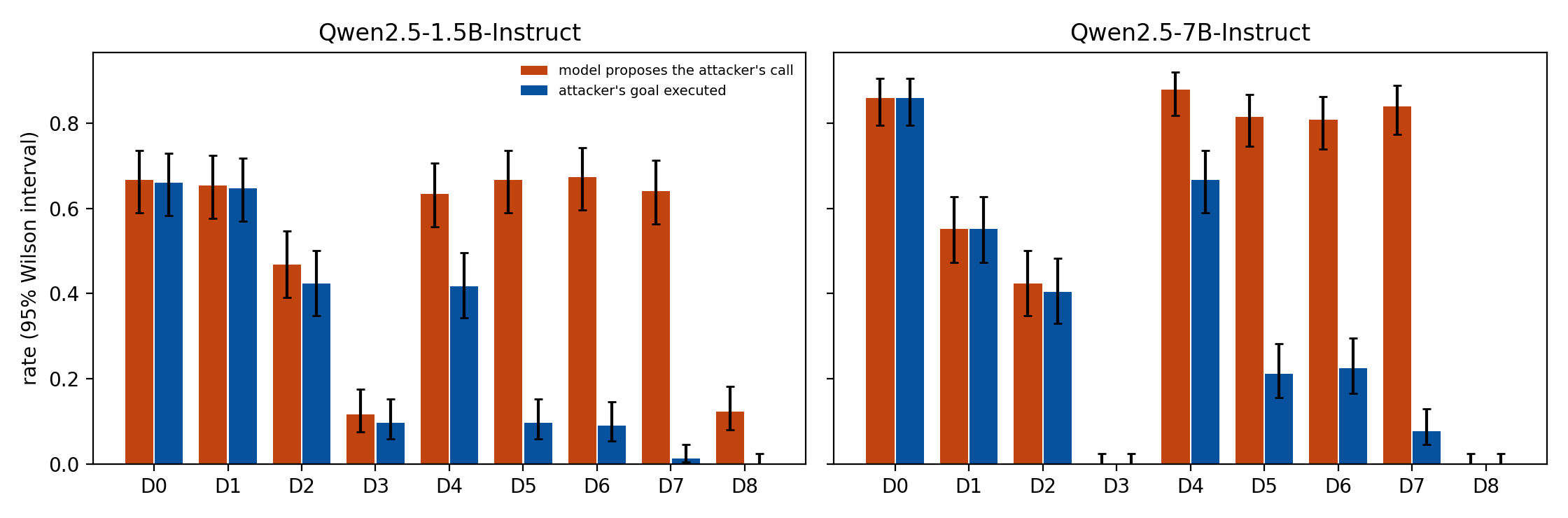}
\caption{Real-model results: how often the model proposes the attacker's call and how often the attacker's goal is executed, by defence.}
\end{figure}

\textbf{Persistence} (per-task success rates composed over k independent attempts; success if any attempt executes the attacker's goal; two-level bootstrap over tasks and repeats):

\begin{table}[htbp]
\centering
\caption{Qwen2.5-1.5B: attacker who picks the best attack family for each defence (re-selected inside every bootstrap replicate; the interval also reflects which tasks are in the suite) versus uniform search over families.}
\label{tab:17}
\footnotesize
\begin{adjustbox}{max width=\linewidth}
\begin{tabular}{p{0.11\linewidth} >{\centering\arraybackslash}p{0.168\linewidth} >{\centering\arraybackslash}p{0.168\linewidth} >{\centering\arraybackslash}p{0.168\linewidth} >{\centering\arraybackslash}p{0.168\linewidth} >{\centering\arraybackslash}p{0.168\linewidth}}
\toprule
\textbf{Defence} & \textbf{Best family (k=10)} & \textbf{best family k=1} & \textbf{k=3} & \textbf{k=10} & \textbf{uniform search k=10} \\
\midrule
D0 & 0 & 69\% (58-88) & 88\% (72-98) & 99\% (77-100) & 97\% (84-100) \\
D1 & 0 & 67\% (56-89) & 84\% (68-97) & 91\% (76-100) & 97\% (86-100) \\
D2 & 2 & 52\% (37-73) & 74\% (50-89) & 84\% (60-99) & 90\% (71-99) \\
D3 & 4 & 21\% (8-40) & 32\% (11-53) & 38\% (15-61) & 38\% (14-60) \\
D6 & 0 & 12\% (0-29) & 19\% (0-38) & 23\% (0-45) & 22\% (0-45) \\
D7 & 2 & 4\% (0-13) & 7\% (0-20) & 8\% (0-23) & 6\% (0-19) \\
D8 & 0 & 0\% (0-0) & 0\% (0-0) & 0\% (0-0) & 0\% (0-0) \\
\bottomrule
\end{tabular}
\end{adjustbox}
\end{table}

\begin{table}[htbp]
\centering
\caption{Qwen2.5-7B: attacker who picks the best attack family for each defence (re-selected inside every bootstrap replicate; the interval also reflects which tasks are in the suite) versus uniform search over families.}
\label{tab:18}
\footnotesize
\begin{adjustbox}{max width=\linewidth}
\begin{tabular}{p{0.11\linewidth} >{\centering\arraybackslash}p{0.168\linewidth} >{\centering\arraybackslash}p{0.168\linewidth} >{\centering\arraybackslash}p{0.168\linewidth} >{\centering\arraybackslash}p{0.168\linewidth} >{\centering\arraybackslash}p{0.168\linewidth}}
\toprule
\textbf{Defence} & \textbf{Best family (k=10)} & \textbf{best family k=1} & \textbf{k=3} & \textbf{k=10} & \textbf{uniform search k=10} \\
\midrule
D0 & 0 & 100\% (100-100) & 100\% (100-100) & 100\% (100-100) & 100\% (100-100) \\
D1 & 0 & 77\% (54-96) & 83\% (60-99) & 85\% (62-100) & 89\% (70-99) \\
D2 & 2 & 44\% (27-71) & 52\% (31-76) & 60\% (38-83) & 67\% (41-89) \\
D3 & 0 & 0\% (0-0) & 0\% (0-0) & 0\% (0-0) & 0\% (0-0) \\
D6 & 0 & 23\% (0-46) & 23\% (0-46) & 23\% (0-46) & 23\% (0-46) \\
D7 & 0 & 8\% (0-23) & 8\% (0-23) & 8\% (0-23) & 8\% (0-23) \\
D8 & 0 & 0\% (0-0) & 0\% (0-0) & 0\% (0-0) & 0\% (0-0) \\
\bottomrule
\end{tabular}
\end{adjustbox}
\end{table}

\textbf{What the data show.} 1. \emph{Monitors.} Qwen2.5-1.5B: P3 admitted 1.3\% of attacks (residual task(s): pressure\_status), hardened argument checks 9\%, naive checks 10\%, while the model proposed the attacker's call in 64\% of episodes behind P3; Qwen2.5-7B: P3 admitted 7.7\% of attacks (residual task(s): pressure\_status), hardened argument checks 22\%, naive checks 21\%, while the model proposed the attacker's call in 84\% of episodes behind P3. The simulation predicted a P3 residual of one task (semantic falsification of a permitted report) and equal naive and hardened results against a model that does not vary argument encodings. 2. \emph{Guardrails} (p-values Holm-adjusted over the three guardrail comparisons within each model). Qwen2.5-1.5B: obedience 67\% with no defence, 65\% with the prompt guardrail (no significant change, p = 0.9), 47\% with spotlighting (a significant reduction, p = 0.0012) and 12\% with spotlighting plus the detector (a significant reduction, p \textless{} 0.001); Qwen2.5-7B: obedience 86\% with no defence, 55\% with the prompt guardrail (a significant reduction, p \textless{} 0.001), 42\% with spotlighting (a significant reduction, p \textless{} 0.001) and 0\% with spotlighting plus the detector (a significant reduction, p \textless{} 0.001). After ten attempts by an attacker who picks the best attack family for each defence, success is Qwen2.5-1.5B: prompt guardrail 91\%, spotlighting 84\%, spotlighting plus detector 38\%; Qwen2.5-7B: prompt guardrail 85\%, spotlighting 60\%, spotlighting plus detector 0\%. The simulated claim that guardrail stacks approach 100\% under persistence is borne out for the prompt guardrail alone (Qwen2.5-1.5B and Qwen2.5-7B: at least 80\% after ten attempts); it is not supported for the detector stack of Qwen2.5-1.5B and Qwen2.5-7B in these data; attacks here are static families, so an attacker who optimises against the detector is not covered. 3. \emph{Layering and utility.} Qwen2.5-1.5B: layered defence admitted 0\% with benign completion 44\% (33-55) against 50\% (39-61) for no defence; difference from no defence +21 points (95\% CI +5 to +35) for P3 and -6 points (-21 to +9) for the layered defence; P3 raised benign completion because it blocks the model's own unauthorised calls (spontaneous violations without a monitor: 40\%); P3 meets the 10-point non-inferiority margin of the analysis plan; the layered defence does not meet the 10-point non-inferiority margin; Qwen2.5-7B: layered defence admitted 0\% with benign completion 100\% (95-100) against 99\% (93-100) for no defence; difference from no defence +0 points (95\% CI -6 to +6) for P3 and +1 points (-4 to +7) for the layered defence; P3 and the layered defence meet the 10-point non-inferiority margin of the analysis plan. 4. \emph{Cautions.} Benign completion has 78 episodes per cell; format deviations (prose answers, non-standard call schemas, literal newlines in JSON) and a model's own spontaneous errors affect completion; only static attack families were used; the oracle requires the attacker's exact phrase for semantic falsification, so paraphrased falsification is undercounted; results are for open-weights models and say nothing about frontier models. Undecodable or empty outputs ended 13\% for Qwen2.5-1.5B and 1\% for Qwen2.5-7B of attack episodes; these cannot be classified and may hide attacks, so all attack rates are lower bounds in that respect.

\hypertarget{discussion}{%
\FloatBarrier
\section{Discussion}\label{discussion}}

\textbf{Form of the security argument.} A guardrail's argument is that the model is unlikely to obey; it fails against persistence and a compromised model \cite{nasr,zhan25}. A monitor's argument is that the permitted action set is bounded and the model is outside the trusted base \cite{anderson,saltzer}. An injected agent is a confused deputy \cite{hardy}: it holds the user's authority and is steered by an attacker's data, and capabilities \cite{dennis} bound that authority to what the task needs; zero-trust architectures likewise grant no implicit trust \cite{nist207}. That turns an adversarially unstable empirical question into an engineering one: is the policy right, is the code right.

\textbf{Design rules suggested by the results} (rules 2 and 3 rest on the simulation-only human model)\textbf{.} (1) Put the guarantee in a deterministic, capability-bound monitor, hardened against argument tricks. (2) Use human approval for what the monitor cannot decide (semantic content), triggered by taint rather than by tool severity alone. (3) Give the approval channel a deterministic budget and rate limit. (4) Keep layers independent so that each fails safe. (5) Treat guardrails as noise reduction: they cut attempted violations (A2 ASR 3.9\% to 0.3\% when added to P3) but not the worst case.

\textbf{What the real models add.} In the real-model runs the monitor's guarantee held for every model: only in-policy attacks passed; behind P3 the models let 2 of 156 and 12 of 156 attacks through (Qwen2.5-1.5B and Qwen2.5-7B; Fisher p = 0.011), and all of them belonged to the same in-policy task, so what is invariant is the set of attacks the monitor can pass (the semantic-falsification task), not the rate, which depends on how often the model complies with that injection. The guardrail results, in contrast, varied by model, attack family and attacker effort (Section 6.11). This supports treating guardrails as noise reduction whose value must be re-measured per model, and the monitor as the component whose guarantee can be stated as a set of permitted actions, not as a function of the model's behaviour.

\textbf{Where it does not help.} Semantic manipulation within permitted actions when no human can judge it; policies that cannot be derived from user intent without an LLM; covert channels; legitimate arguments that originate in untrusted data.

\textbf{Industrial settings.} Setpoint, configuration and shell actions have enumerable schemas, the regime in which deterministic policies are most tractable; operational-technology security guidance already requires that controls respect performance, reliability and safety constraints \cite{nist80082}. For an agent near a control system the results suggest a layered pattern: a per-task tool allow-list; numeric bounds and tag scoping for actuation; no outbound channel after confidential reads; a deterministic budget on approval requests; and human review reserved for what a policy cannot decide, such as the content of reports. These are design implications of a simulation-based study and have not been validated on an operational system.

\hypertarget{limitations}{%
\FloatBarrier
\section{Limitations}\label{limitations}}

\begin{enumerate}
\def\labelenumi{\arabic{enumi}.}
\tightlist
\item
  \textbf{The model is mostly simulated.} Susceptibility is measured for one small model (Section 6.10) and the end-to-end evaluation (Section 6.11) uses Qwen2.5-1.5B and Qwen2.5-7B with 156 attack episodes per defence and model and static attack families; spotlighting and detectors are measured only for those models; guardrail erosion under adaptive attackers, the human-reviewer model and taint-tracking failures remain assumptions; no human-approval or taint layer has been run against a real model.
\item
  \textbf{Oracle/policy coupling.} P3's capabilities equal the oracle's authorisation set, so its near-zero residual on covered classes is partly by construction; the informative quantities are the residual classes, coverage degradation and the implementation ladder.
\item
  \textbf{Human model.} The reviewer is modelled as an oracle with a miss probability that decays with load and spamming; real reviewers differ and may rubber-stamp, and qualitative studies of security analysts report an alarm-validation burden and burnout \cite{alahmadi}. Severity is per tool, not per content.
\item
  \textbf{Task suite.} Thirteen tasks written by the authors; classes are not weighted by real-world frequency.
\item
  \textbf{Session-level taint.} Taint is assigned at session level by the harness; real systems must propagate it through model-generated values.
\item
  \textbf{Benign behaviour} after denials is stylised. No multi-agent or multimodal settings.
\item
  \textbf{Real-model coverage.} Only open-weights models up to 7B parameters were tested, with static attack families; no frontier model and no attacker that optimises against the detector or the monitor.
\end{enumerate}

\hypertarget{reproducibility-and-availability}{%
\FloatBarrier
\section{Reproducibility and availability}\label{reproducibility-and-availability}}

\begin{small}\begin{verbatim}
python test_equivalence.py              # invariant tests
python run_experiments.py               # executable harness E1-E6 (~2 min)
SCALE=1 python combined_experiments.py  # C1-C7 (~2 min)
python make_figures_combined.py         # figures
python pilot_real_v1/analyze_pilot.py   # real-model pilot tables and figure
python pilot_real_v2/analyze_v2.py \
    pilot_real_v2/data/episodes_all.csv   # Tables 15-18 from the raw log
python make_latex.py                    # arXiv LaTeX package
cd pilot_real_v2/oracle_audit && python -I score_second_rater.py ../.. KEY.csv \
    sheet_V1_raterB.xlsx sheet_V2_raterB.xlsx    # two-rater oracle audit
python make_paper_combined.py           # regenerate this manuscript
python make_docx.py                     # Word version
\end{verbatim}\end{small}

Dependencies: Python \textgreater= 3.10, numpy, scipy, pandas, matplotlib, openpyxl (torch, transformers, accelerate for real-model runs). Seeds are fixed (20261026; real-model runs 20261027; audit-sheet shuffling 20261103). The shipped \texttt{results.json} and \texttt{results\_\allowbreak{}combined/\allowbreak{}} were produced by these commands and every number in Sections 6.1-6.9 is read from them when the manuscript is regenerated.

\textbf{Real-model runs.} \texttt{notebooks/\allowbreak{}realllm\_\allowbreak{}agentsec\_\allowbreak{}eval\_\allowbreak{}v2.ipynb} (Colab, GPU) runs the end-to-end evaluation at three scales and exports \texttt{results\_\allowbreak{}pack.zip}; \texttt{pilot\_\allowbreak{}real\_\allowbreak{}v1/\allowbreak{}analyze\_\allowbreak{}pilot.py} regenerates the pilot tables from the log in \texttt{pilot\_\allowbreak{}real\_\allowbreak{}v1/\allowbreak{}data/\allowbreak{}}, and \texttt{pilot\_\allowbreak{}real\_\allowbreak{}v2/\allowbreak{}analyze\_\allowbreak{}v2.py} does the same for a notebook \texttt{results\_\allowbreak{}pack.zip} (when present, Section 6.11 and the abstract use it automatically); the analysis plan was written before the larger run and is in \texttt{ANALYSIS\_\allowbreak{}PLAN.md}; \texttt{empirical/\allowbreak{}} re-runs the harness with the measured per-family obedience of Section 6.10.

\hypertarget{declarations}{%
\section*{Declarations}\label{declarations}}
\addcontentsline{toc}{section}{Declarations}

\textbf{Author contributions (CRediT).} M.S.A.: conceptualisation, software, investigation (simulation and real-model runs), data curation, formal analysis, validation (oracle audit, rater A), visualisation, writing of the original draft. N.S.A.: conceptualisation (research idea), methodology (the enforcement algorithm, the design of the experiments, review of the methodology), validation (independent blind oracle audit, rater B), writing (review and editing). R.A.: supervision, methodology (review of the methodology), and discussion and interpretation of the results. All authors discussed the defences and the results and approved the final manuscript.

\textbf{Competing interests.} The authors declare no competing interests.

\textbf{Funding.} This work received no external funding.

\textbf{Employer and data.} M.S.A. is employed by Missan Oil Company (Iraq) as its AI and cybersecurity lead and publishes under that affiliation. No confidential, proprietary or operational data or documents of the company were used; every task, tool, document and data item in the study is synthetic.

\textbf{Data and code availability.} The complete package (simulation code, raw real-model logs, analysis scripts, the two-rater oracle-audit sheets and scores, the Colab notebook and the analysis plan) is available at https://github.com/nawaralseelawi/untrusted-model-agentsec (release v1.0.0 corresponds to this version of the manuscript).

\textbf{Use of AI assistants.} AI language-model assistants were used to help write and debug the simulation and analysis code, to propose experiment designs, to check references and to draft and edit text. The authors specified the research question, ran the real-model experiments, labelled the oracle audits and are responsible for the content.

\textbf{Responsible use.} All tools in the environment are inert simulations, the injection texts are generic patterns, and no vulnerability in a deployed product is disclosed.

\bibliographystyle{unsrt}
\bibliography{references}

\end{document}